\documentclass[conference]{IEEEtran}
\IEEEoverridecommandlockouts
\usepackage{cite}
\usepackage{amsmath,amssymb,amsfonts}
\usepackage{algorithmic}
\usepackage{graphicx}
\usepackage{textcomp}
\usepackage{subcaption}
\usepackage{float}
\usepackage{tikz}
\usepackage[T1]{fontenc}
\usepackage{times}
\usepackage{mathptmx}
\usepackage[scaled=0.92]{helvet}
\usepackage{courier}
\usepackage{url}
\usepackage{balance}

\usetikzlibrary{shapes.geometric, shapes.misc, arrows.meta}
\usepackage{xcolor}

\def\BibTeX{{\rm B\kern-.05em{\sc i\kern-.025em b}\kern-.08em
    T\kern-.1667em\lower.7ex\hbox{E}\kern-.125emX}}
\begin{document}
\title{Quantifying System Level KPI Deviations of Sionna RT: Material and Near-Field Error Analysis Using a 5G OAI Testbed

\thanks{This work was supported in part by the German Federal Ministry of Research Technology and Space (BMFTR) in the course of the 6GEM+ Transfer Hub under grant 16KIS2411}
}

\author{\IEEEauthorblockN{Faizan Rauf, Srijita Sanyal, 
Markus Heinrichs and Aydin Sezgin}
\IEEEauthorblockA{
\text{Ruhr-Universität Bochum}, Bochum, Germany\\
\{faizan.rauf, srijita.sanyal, markus.heinrichs, aydin.sezgin\}@rub.de}}

\maketitle

\begin{abstract}

Ray tracing (RT) has recently gained renewed interest in wireless communications, driven by its integration into digital twin (DT) frameworks for site specific channel modeling. Several previous studies have validated RT at the channel level, yet how these errors propagate into real 5G system level key performance indicators (KPIs) on actual hardware remains unquantified. This paper addresses this gap by comparing Sionna RT simulated channels against vector network analyzer (VNA) measured channels using an OpenAirInterface (OAI) 5G NR testbed. Channel measurements are conducted at 20 receiver positions in an indoor laboratory, with both channel types injected into a hardware in the loop channel emulator interfacing an OAIBOX MAX base station and a Quectel UE. RSRP, PUCCH SNR, and SINR are evaluated under both conditions. The results identify antenna near-field transition effects as a critical position-dependent error source, alongside material property mismatch, providing a quantitative benchmark for digital twin-based 5G and beyond network planning.

\end{abstract}

\section{Introduction}
Ray Tracing (RT) has lately gained wide recognition in the simulation of wireless communications \cite{7152831}, for the prediction of channels in given environments. The advantages of RT compared to traditional simulation tools lie in the consideration of the geometric and material properties of the scene \cite{hoydis2023sionna}. With information about the inherent properties that make up the scene, the electromagnetic properties of wave propagation become more precisely predictable, therefore helping the visualization of the channel properties. 
However, the end goal of any research is seldom the prediction of just the channel but the extended use of said determined channel and conditions, which are then used in further applications. In many such critical applications, the margin of tolerable error is very low, therefore, requiring an understanding of how accurate the predicted channel conditions are and what margin of error the simulated data has to consider, in order to implement accurate compensation technologies. 

Ray tracing has been widely studied as a channel modeling 
tool, with validation against real-world measurements 
forming a central research thread. In~\cite{karstensen2016comparison}, RT simulations were compared 
with vector network analyzer (VNA)-based measurements at 26--30~GHz 
in indoor environments, showing strong agreement for dominant 
propagation paths, though richer multipath components observed 
in measurements were not fully captured by the simulation.
In~\cite{di2024validation,di2025ray}, a home-developed RT 
tool was validated for mmWave massive MIMO systems using a virtual uniform circular array (UCA), showing that third-order reflections 
sufficiently capture dominant indoor paths, with LoS 
deviations below 2~dB. At sub-THz frequencies, 
\cite{dupleich2024measurement} validated Sionna RT against 
measurements at 187.5~GHz in an industrial scenario using 
a LiDAR-based digital twin, confirming geometric agreement 
while identifying model simplification as a key error 
source. The work in~\cite{zhu2025digital} extended 
DT-based RT validation to 300~GHz for in-vehicle V2X 
scenarios, further leveraging the validated model for 
system-level wireless planning. At the network level, 
\cite{manukyan2025limitations} evaluated Sionna RT fidelity 
in an outdoor urban setting, showing that antenna placement 
and orientation dominate simulation accuracy over solver 
parameters.

The prediction errors identified in these studies 
have motivated a growing body of work on digital 
twin calibration \cite{hoydis2023sionna, ruah2024calibrating, 
luo2026wireless, zhao2024learnable, 
vlm2025differentiable}. In \cite{hoydis2023sionna}, the 
differentiable nature of Sionna RT is exploited to 
optimize material electromagnetic properties via 
gradient descent, directly minimizing the error 
between predicted and measured channel responses. 
Building on this, \cite{ruah2024calibrating} 
proposed a calibration framework that corrects 
local phase errors in RT predictions using 
measurement-based estimates, improving geometric 
accuracy of the digital twin. More recently, 
\cite{luo2026wireless} introduced a deep learning 
approach that refines the DFT-domain channel 
information generated by a low-complexity digital 
twin using CSI feedback, bridging the gap between 
synthetic and real channels without rebuilding the 
digital twin model itself. While these methods 
offer promising calibration strategies, they 
predominantly evaluate performance at the channel 
coefficient level. However, how RT channel errors propagate into system level KPIs 
through a real 5G testbed, and which modeling aspects 
drive these deviations, remains largely unaddressed.

Closest to our work,~\cite{iye2025open} proposed an Open Wireless 
Digital Twin (OWDT) platform integrating OpenAirInterface (OAI) 
and Sionna RT for end-to-end 5G mobility emulation in an open 
radio access network (O-RAN) framework, demonstrating real-time 
KPI monitoring including reference signal received power (RSRP), 
modulation and coding scheme (MCS), block error rate (BLER), and 
throughput under vehicular mobility scenarios in urban environments.
However, their work relies solely on Sionna RT-generated 
channels without validating these against physically 
measured channels. The deviation between RT-simulated 
and real measured channels at the system KPI level 
therefore remains unquantified.

This paper addresses this gap by directly 
comparing Sionna RT channels against VNA measured 
channels in an indoor scenario using an OAI based 
5G testbed with a channel emulator, enabling a 
quantitative assessment of system level KPI 
deviations introduced by RT channel modeling 
errors. To the best of our knowledge, this 
hardware validated system level analysis has 
received limited attention in the existing 
literature, and the findings directly inform 
the design of more effective digital twin 
calibration strategies.

 This paper is organized as follows. 
Section~\ref{sec:indoorscene} describes the indoor 
laboratory environment, detailing both the 
digital twin construction in Blender, the 
Sionna RT simulation configuration, and the 
VNA based channel measurement setup. 
Section~\ref{sec:Methodology} presents the 
system methodology, covering the OAI-based 
5G NR testbed architecture and the 
hardware-in-the-loop channel emulator 
configuration used for end-to-end KPI 
evaluation. 
Section~\ref{sec:results} presents the experimental results, 
quantifying RT prediction errors across RSRP, physical uplink control channel (PUCCH) signal to noise ratio (SNR), and signal to interference plus noise 
ratio (SINR) metrics at 20 spatially distributed receiver positions 
within the indoor environment. Finally, Section~\ref{sec:conclusion} 
concludes the paper and outlines directions for future work.

\section{Indoor Scene Setup}
\label{sec:indoorscene}
In this section, we discuss how the scene is set up for simulation and measurements. The environment is an indoor lab at the department of Digital Communication Systems, Ruhr-Universität Bochum. The lab is about $3.45\mathrm{m} \times 5.4\mathrm{m}$  in dimension and the ceiling is $3.66\mathrm{m}$ high. The walls and floor are made of concrete and the ceilings are made of ceiling boards. There is a long stretch of windows with metal frames on one end of the room, and a door made of wood on the other end, with a small glass look-through on top. Underneath the window stretch, there is a marble base, which tops the electrical insulation box made of Polyvinyl Chloride (PVC). On the upper ends of two adjacent walls, there are encasing made of plasterboard. There is a large wooden table with metal legs placed in the center of the room. On either side of the room, there are open-faced cupboards made of plywood. The shelves in these cupboards are also made of plywood. There are closed plywood cabinets on either side of the door, but their shelves are made of metal. The cabinet on the left side of the door has metal poles stored inside.The R\&S ZNA is housed in a rigid steel chassis with 
aluminum alloy front and rear panels, making it a 
significant metallic object within the scene. In the 
Sionna RT simulation, it has been assigned metallic 
material properties, as steel and aluminum exhibit high 
electrical conductivity and act as strong specular 
reflectors at microwave frequencies. The diversity of materials present in this environment including concrete, 
wood, metal, glass, plasterboard, and PVC makes it a particularly rich 
and challenging testbed for evaluating the sensitivity of RT-based digital 
twins to material property assignments.

\subsection{Simulation Setup}\label{AA} 
The Digital Twin of the indoor scene, as shown Fig. 1, has been created in Blender v4.2, with attention to  detail. The ray tracing simulation is conducted in Sionna RT v1.2.1.
\begin{figure}[t]
    \centering
    \includegraphics[width=1\linewidth, height=12cm, 
    keepaspectratio=false]{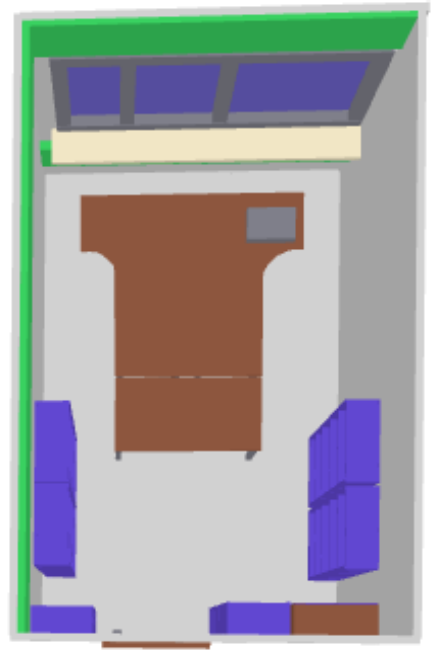}
    \caption{Top-view of the Digital Twin scene created in 
Blender v4.2, showing the indoor lab layout including 
the wooden desk, shelving units, window, and door used 
in the Sionna RT simulation.}
    \label{fig:dt1}
\end{figure}
All the objects in the scene have been assigned their accurate material and geometric properties. For simplicity, PVC has been assigned to the closest predefined ITU material in Sionna, which is plasterboard. The ray tracer has been set with a maximum depth of 6, allowing at most rays that have been reflected six times. This is to limit both the computation load and paths with high delay but also reserve the multipath information.. We allow Line-of-Sight (LoS) paths and specular reflections. For now, we keep the diffused reflections and refractions turned off, since the most substantial part of the glass is one one end of the room and the antennas are turned away from the direction. The transmitter and receiver are each configured as a single 
antenna element following the antenna pattern defined in 
3GPP TR\,38.901~\cite{3gpp_tr38901}. We simulate the channel impulse responses (CIR), Channel Frequency
Response (CFR), and the discrete channel taps (DCTs). For the CIRs, we clip the output at a maximum of 20 paths.

\subsection{Measurement Setup}
\begin{figure}[!b]
    \centering
    \includegraphics[width=1\linewidth, height=10cm, 
    keepaspectratio=false]{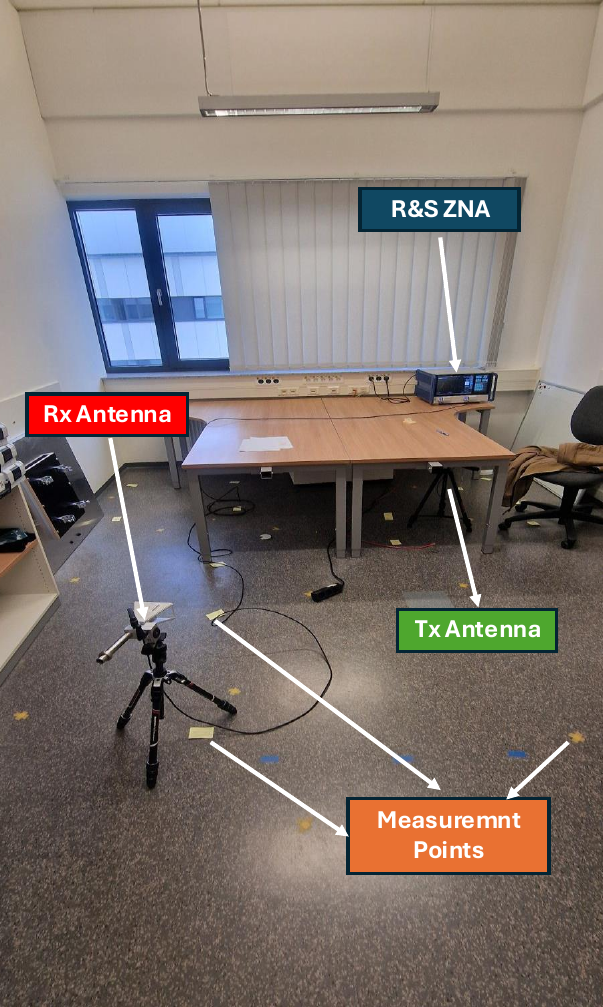}
    \caption{Indoor lab measurement setup showing the Tx and Rx
    horn antennas, R\&S ZNA vector network analyzer, and UE
    measurement grid positions on the floor.}
    \label{fig:meas1}
\end{figure}

The channel measurements were carried out using a VNA, specifically the Rohde \& Schwarz ZNA \cite{rs_zna},
as illustrated in Fig.~\ref{fig:meas1}. The instrument provides
two measurement ports, denoted as Port~1 and Port~2, which were
used as the transmitter (Tx) and receiver (Rx) ports,
respectively. The Schwarzbeck BBHA~9120L double ridge broadband 
horn antennas \cite{schwarzbeck_bbha9120l} were connected to Port~1 and Port~2 to serve 
as the transmit and receive elements of the wireless link, 
and were interfaced to the VNA via coaxial SMA cables. These antennas operate over a frequency range of
3\,GHz--40\,GHz, making them well suited for measurements in
the 5G NR n78 band.

The measurement parameters are summarized in
Table~\ref{tab:vna_params}. The center frequency and bandwidth
were selected in accordance with the 5G NR band n78
(3.3--3.8\,GHz), which is widely deployed for sub-6\,GHz 5G
networks. Specifically, a center frequency of 3.349\,GHz with
a bandwidth of 100\,MHz was chosen to align with the channel
bandwidth supported by the 5G NR standard for this band,
ensuring that the measured channel characteristics are
representative of real 5G NR operating conditions.

Prior to the measurements, a full two-port calibration was
performed to eliminate the effect of cable losses, connector
mismatches, and phase delays introduced by the measurement
cables. For this purpose, a Keysight calibration kit was
employed, which facilitated a systematic Short-Open-Load-Thru
(SOLT) calibration procedure. The calibration was carried out
independently on Port~1 and Port~2, where each port was
successively terminated with the Open, Short,
and Load standards of the calibration kit.
Subsequently, the Thru calibration was performed by
directly interconnecting Port~1 and Port~2, establishing a
through connection between the two ports. This procedure
ensured full error correction across all measurement ports,
such that the measured S\textsubscript{21} transmission
response represents only the wireless propagation channel
between the antennas, free from any instrumentation artifacts.

Measurements were conducted within the indoor room environment described in Section~\ref{sec:indoorscene}. As shown in
Fig.~\ref{fig:meas1}, the Tx antenna was kept fixed at a
predefined position, while the Rx antenna was successively
moved to 20 different receiver positions distributed throughout
the measurement area. At each position, the CFR was recorded over the selected bandwidth using
the VNA, yielding a set of 20 spatially distributed channel
measurements for subsequent comparison with the Sionna RT
simulations. The 1001 frequency points across 100\,MHz
correspond to a frequency resolution of 100\,kHz, which
provides a maximum unambiguous delay of 10\,\textmu s.
This is sufficient to capture all relevant multipath components in
the indoor environment.
\begin{table}[h!]
\centering
\caption{VNA Measurement Parameters}
\label{tab:vna_params}
\begin{tabular}{lc}
\hline
\textbf{Parameter}       & \textbf{Value}              \\
\hline
Center Frequency         & 3.349\,GHz                  \\
Frequency Range          & 3.299--3.399\,GHz           \\
Bandwidth                & 100\,MHz                    \\
Frequency Points         & 1001                        \\
Frequency Resolution     & 100\,kHz                    \\
Max. Unambiguous Delay   & 10\,\textmu s               \\
Antenna Type             & Schwarzbeck BBHA~9120L      \\
Antenna Frequency Range  & 3\,GHz--40\,GHz           \\
5G NR Band               & n78                         \\
Number of Rx Positions   & 20                          \\
Tx Position              & Fixed                       \\
Calibration Method       & SOLT (2-port)               \\
\hline
\end{tabular}
\end{table}

\section{Methodology}
\label{sec:Methodology}
\begin{figure}[h!]
    \centering
    \includegraphics[width=1\linewidth, height=8cm, 
    keepaspectratio=false]{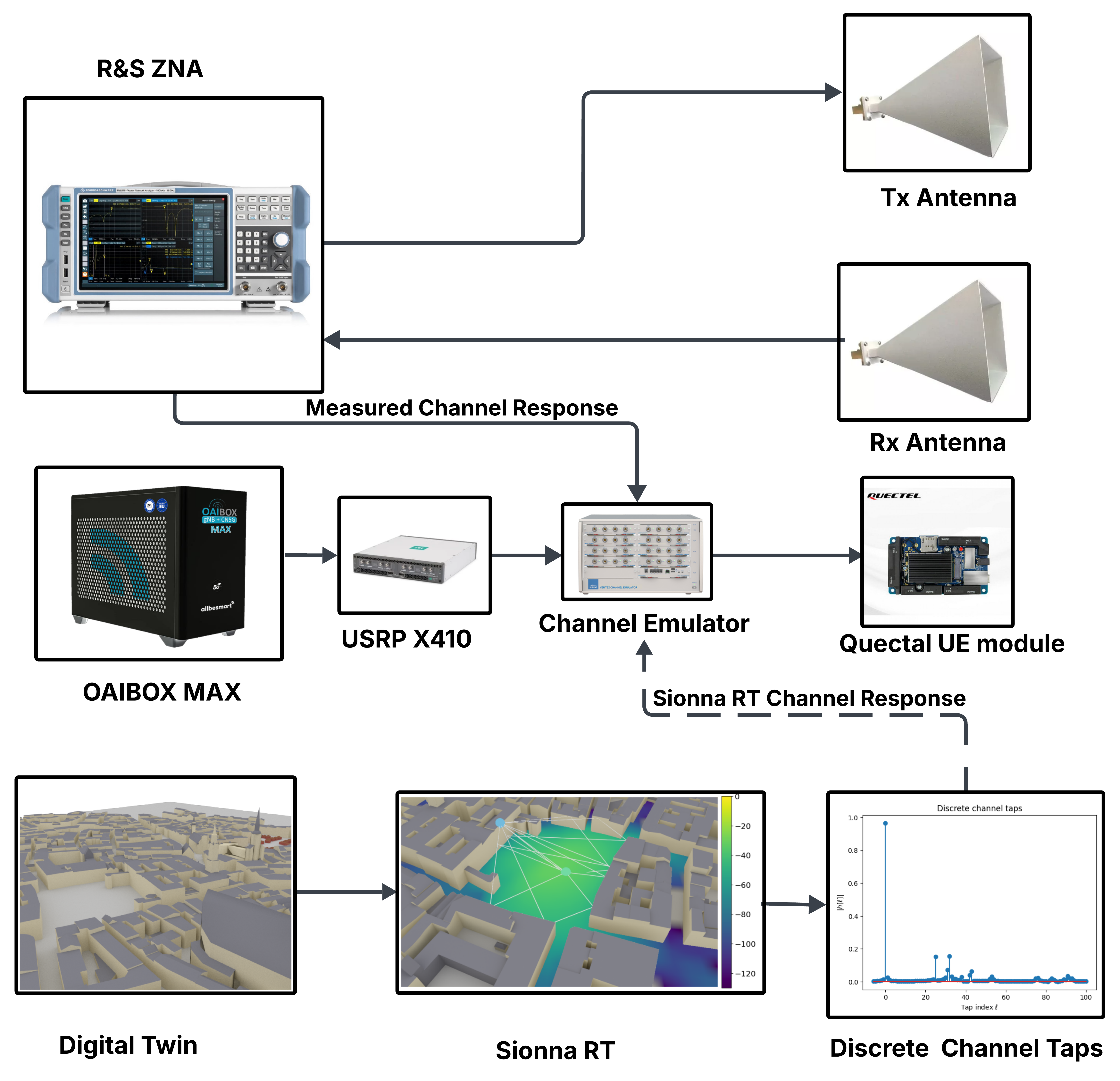}
    \caption{Proposed hardware-in-the-loop testbed integrating 
VNA measured and Sionna RT simulated channels for end-to-end 
5G NR KPI evaluation.}
    \label{fig:blockdiagram}
\end{figure}

The proposed testbed enables a direct comparison between 
physically measured wireless channels and Sionna RT-simulated 
channels at the system KPI level. As illustrated in 
Fig.~\ref{fig:blockdiagram}, the channel emulator serves 
as the central element of the testbed, receiving discrete 
channel taps from either the VNA measurement chain or the 
Sionna RT simulation pipeline described in 
Section~\ref{sec:indoorscene}, and applying them to the 
live 5G NR signal in real time.
\subsection{5G NR Testbed}

\begin{figure}[h!]
    \centering
    \includegraphics[width=1\linewidth, height=5cm, 
    keepaspectratio=false]{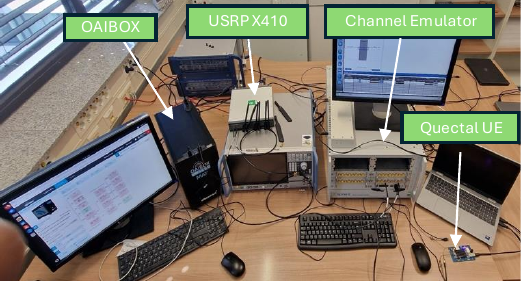}
   \caption{Physical hardware testbed used for end to end 
5G NR KPI evaluation.}
\label{fig:hardware}
\end{figure}

The 5G NR base station is implemented on the 
OAIBOX MAX, a high-performance compute platform 
running the OAI 5G NR protocol stack as shown in Fig.~\ref{fig:hardware}. 
The OAIBOX MAX is configured to operate on the 5G NR 
band n78 (3.3--3.8\,GHz) with a channel bandwidth of 
100\,MHz. The baseband signal generated by the OAI stack 
is passed to the NI USRP X410 software defined 
radio (SDR), which serves as the RF front-end and performs 
digital-to-analog conversion and upconversion at the 
configured center frequency of 3.349\,GHz. The transmitted 
signal is then routed through the channel emulator, which 
applies the configured channel response before delivering 
the signal to the Quectel UE module. The Quectel 
module acts as the 5G NR user equipment (UE) and 
establishes a live radio link with the OAIBOX MAX. 
KPIs including downlink (DL) 
and uplink (UL) maximum bitrate,  SNR, SINR and RSRP are recorded at each Rx position.
\subsection{Channel Emulator}

The channel emulator applies a configurable CIR to the transmitted 5G NR baseband 
signal in real time, replicating the effect of the 
wireless propagation environment without requiring 
over the air transmission. It is configured to match 
the 100\,MHz system bandwidth, ensuring consistency 
with both the VNA measurement setup and the OAI 
testbed configuration. The emulator is operated in 
two modes: first using the VNA-measured discrete 
channel taps as the ground truth reference, and 
subsequently using the Sionna RT-simulated channel 
taps for the same Rx positions. This substitution, 
under otherwise identical testbed conditions, enables 
a direct and quantitative assessment of the KPI 
deviations introduced by RT channel modeling errors.

\section{Results and Discussion}
\label{sec:results}

This section presents the simulation and experimental results obtained by comparing 5G NR KPIs recorded under VNA-measured channels against those obtained under Sionna RT-simulated channels across 20 Rx positions in the indoor environment. Prior to the KPI comparison, the near-field region of the transmit antenna is characterized through CST simulation.

\subsection{Near-Field Characterization}

\begin{figure}[h!]
    \centering
    \includegraphics[width=0.9\linewidth]{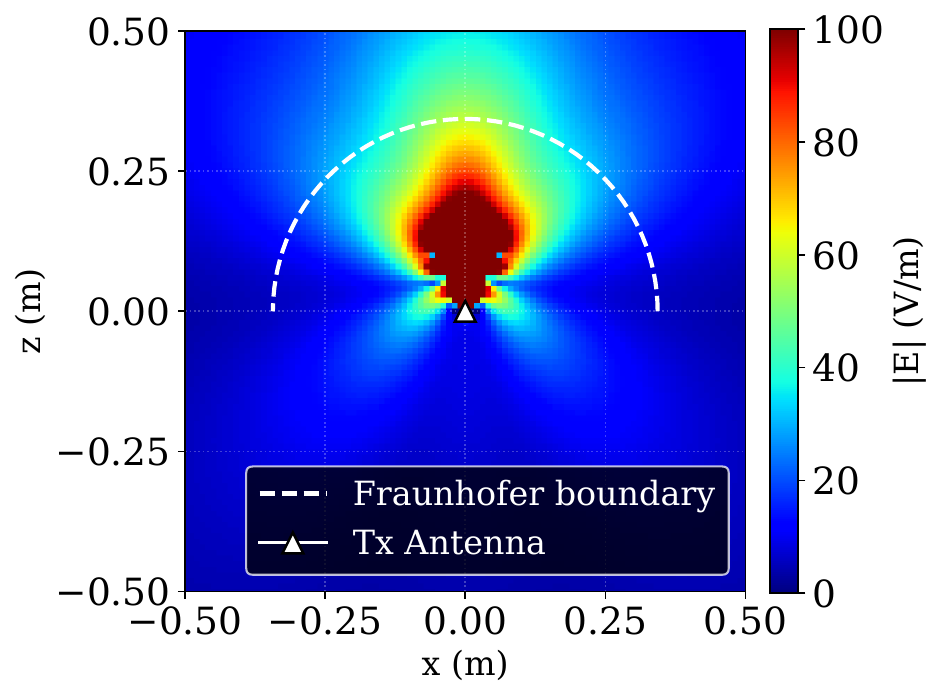}
    \caption{E-field magnitude of the Schwarzbeck BBHA 9120L at 3.349\,GHz. White dashed semicircle marks the Fraunhofer boundary $R_0 = 0.343$\,m.}

    \label{fig:nearfield}
\end{figure}

Since the antenna datasheet provides no radiation pattern data in the 
3--4\,GHz band, the near-field boundary was first determined analytically 
using the Fraunhofer distance criterion:
\begin{equation}
    R_0 = \frac{2D^2}{\lambda} = 
    \frac{2 \times (0.124)^2}{0.0896} \approx 0.343\,\text{m}
    \label{eq:fraunhofer}
\end{equation}
where $D = 124$\,mm is the antenna aperture diameter and 
$\lambda = 89.6$\,mm is the free-space wavelength at 3.349\,GHz. 
To validate this result, the antenna was simulated in CST Microwave 
Studio at 3.349\,GHz over a $3 \times 3$\,m observation plane in 
the XZ plane. The Fraunhofer boundary corresponds to the distance 
at which the maximum phase deviation across the aperture equals 
$\lambda/16$, i.e., 22.5\textdegree. The simulated E-field magnitude distribution, shown in 
Fig.~\ref{fig:nearfield}, confirms that the near-field region 
is confined within $R_0 = 0.343$\,m, in agreement with the 
analytical result of~\eqref{eq:fraunhofer}.

\begin{figure*}[t]
    \centering
    \begin{subfigure}[b]{0.32\linewidth}
        \includegraphics[width=\linewidth]
        {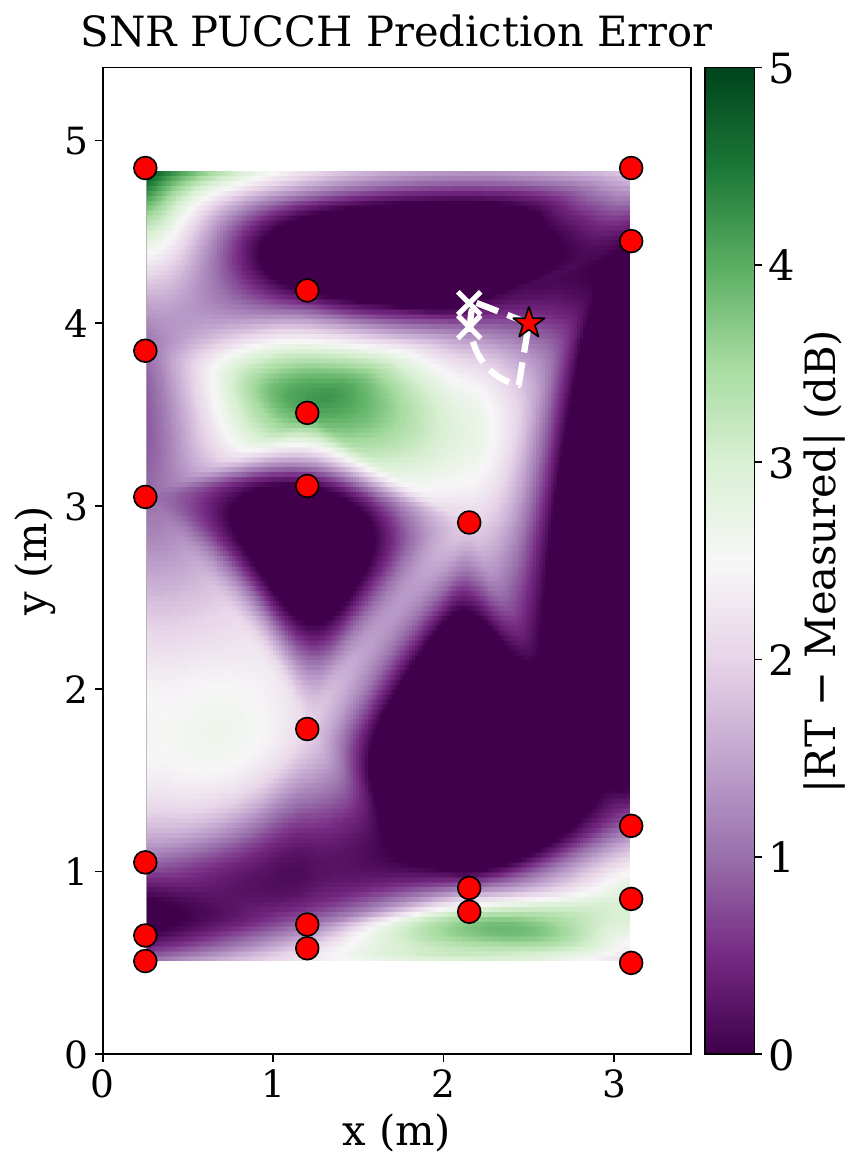}
        \caption{SNR PUCCH Error}
        \label{fig:pusch}
    \end{subfigure}
    \hfill
    \begin{subfigure}[b]{0.32\linewidth}
        \includegraphics[width=\linewidth]
        {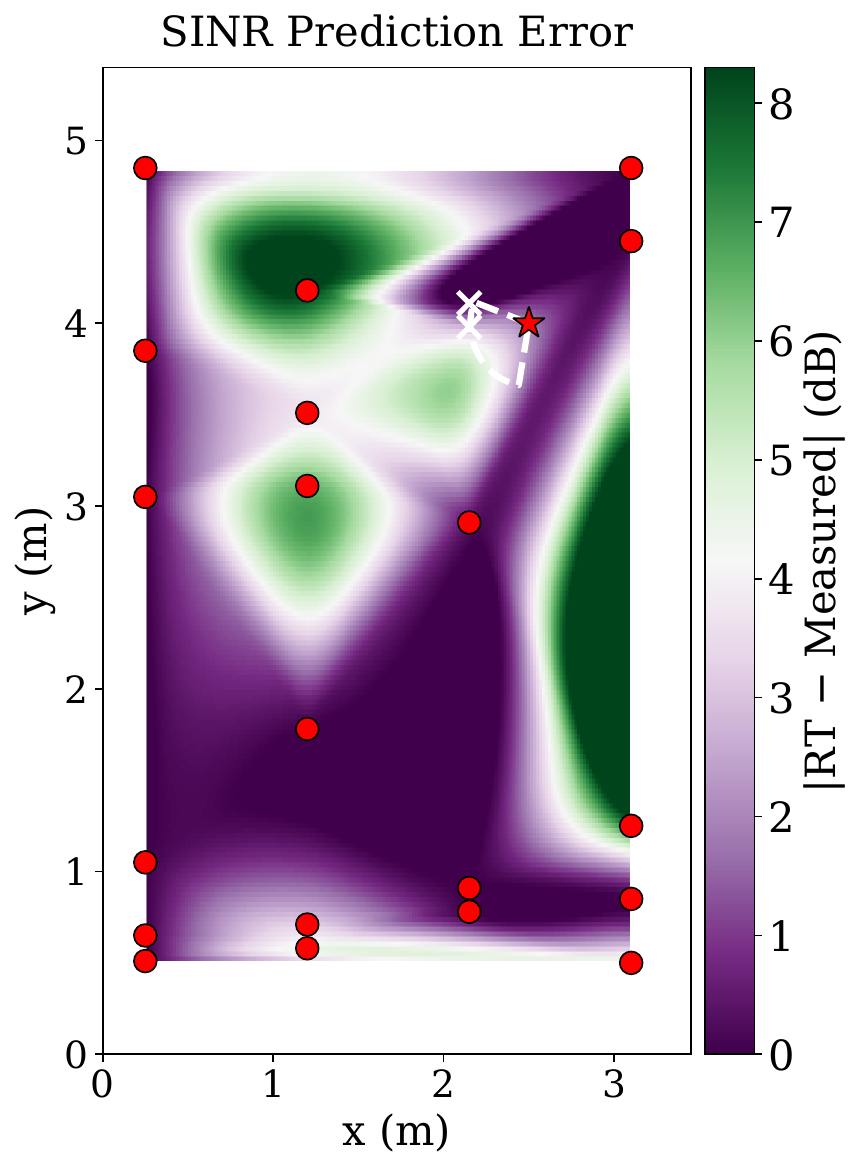}
        \caption{SINR Error}
        \label{fig:pucch}
    \end{subfigure}
    \hfill
    \begin{subfigure}[b]{0.32\linewidth}
        \includegraphics[width=\linewidth]
        {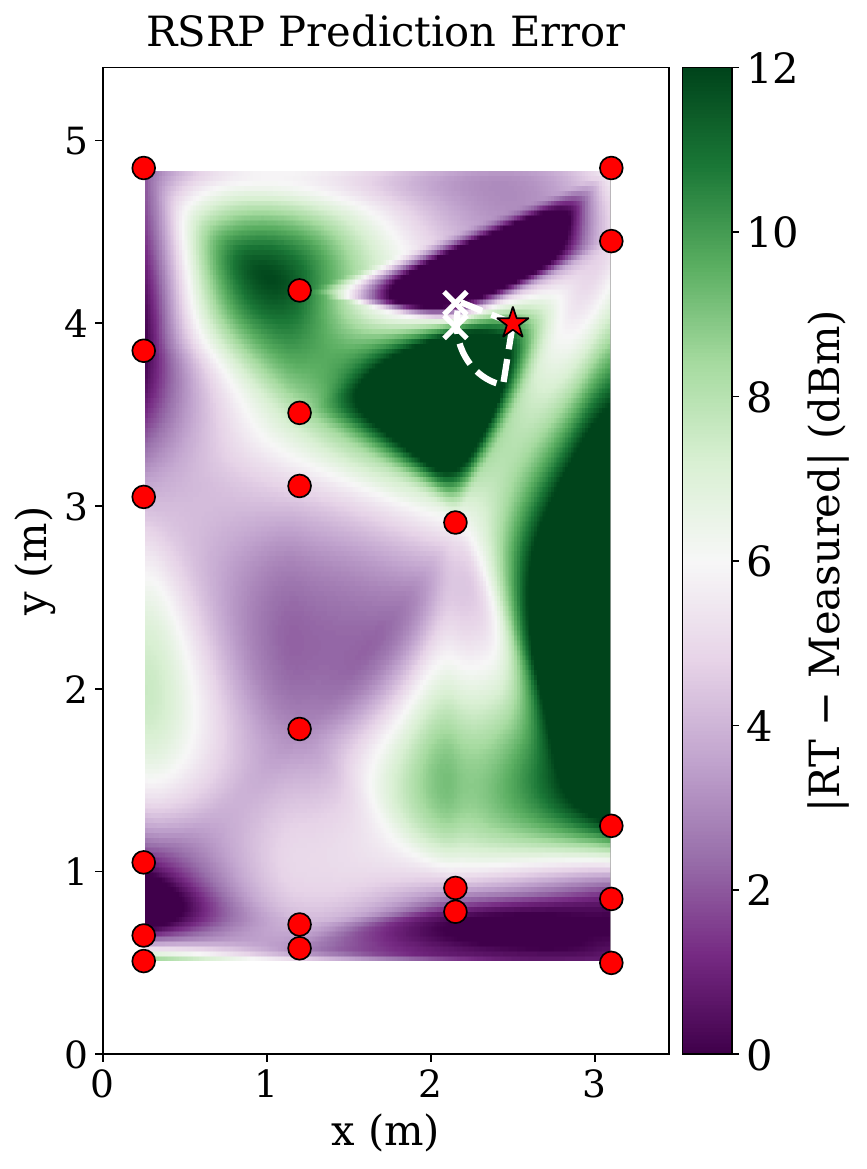}
        \caption{RSRP Prediction Error}
        \label{fig:rsrp}
    \end{subfigure}
     \vspace{4pt}
\begin{center}
\begin{tikzpicture}[font=\small]

    
    \node[star, star points=5, fill=red, draw=black,
          minimum size=8pt, inner sep=0pt] at (0,0) {};
    \node[anchor=west] at (0.2,0) {Tx Antenna};

    \draw[red, dashed, thick] (2.4,0) -- (3.1,0);
    \node[anchor=west] at (3.15,0) {Near-field boundary ($R_0$)};


    \filldraw[fill=red, draw=black] (0.1,-0.5) circle (3.5pt);
    \node[anchor=west] at (0.3,-0.5) {Rx positions};

    
    \draw[black, thick] (2.4,-0.62) -- (2.8,-0.38);
    \draw[black, thick] (2.4,-0.38) -- (2.8,-0.62);
    \node[anchor=west] at (2.9,-0.5) {Near-boundary Rx (Pos. 11, 16)};

\end{tikzpicture}
\end{center}
    \caption{Spatial distribution of absolute prediction errors 
$|$RT $-$ Measured$|$ for (a) SNR PUCCH, (b) SINR, and (c) RSRP 
across 20 receiver positions in the indoor laboratory environment. 
The red star marks the transmit antenna, the white dashed arc 
indicates the Fraunhofer near-field boundary ($R_0 = 0.343$\,m), 
white $\times$ symbols denote positions within the 
near-field transition zone (Pos. 11 and 16) and red circle shows different Rx position.}
    \label{fig:heatmaps}
\end{figure*}
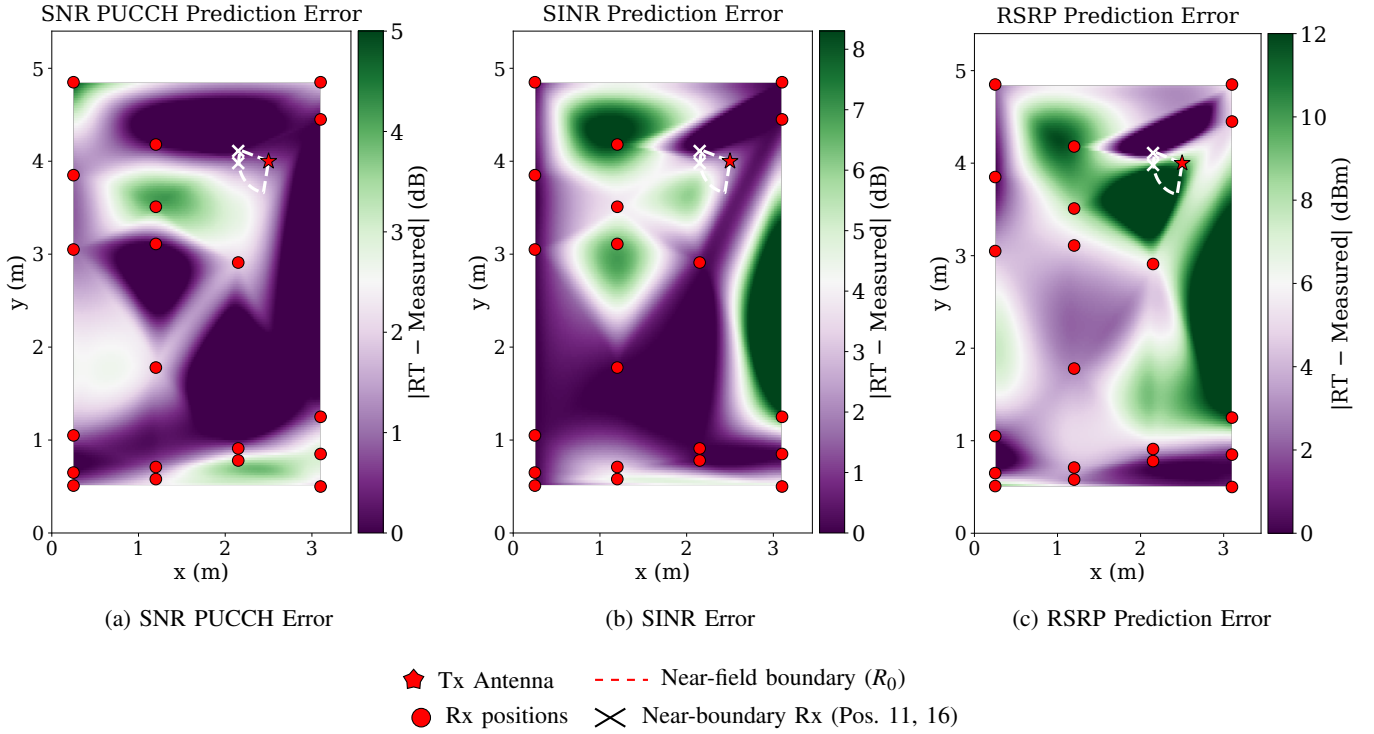
\subsection{Spatial Error Analysis}
Figs.~\ref{fig:pusch}, \ref{fig:pucch}, 
and~\ref{fig:rsrp} present the spatial distribution 
of the SNR PUCCH, SINR and RSRP prediction 
errors (RT $-$ Measured) across the indoor 
measurement area. Three distinct error regions 
are identified.

A first source of systematic error is observed in regions
surrounding the furniture, particularly near the cupboards and
under the table. The prediction error is notably high in these areas
across all three heatmaps, which is primarily attributed to
material property mismatch in the digital twin. The cupboards
are constructed from lighter and thinner wood than conventional
furniture, yet are classified as standard wood in the Sionna~RT
scene for implementation convenience. Similarly, the table,
while classified as wood, has metal legs whose reflective
properties differ significantly from the assigned material
parameters. Furthermore, the PVC encasings of the electrical
wiring along the back wall are approximated as plasterboard in
the digital twin, as this represents the closest available
material in terms of electromagnetic parameters. These
approximations introduce localized multipath prediction errors,
particularly in regions where reflected and scattered rays from
these surfaces contribute significantly to the received signal.
This underscores the critical importance of precise material
calibration in indoor RT-based digital twins, where dense
multipath environments amplify even small deviations in
material parameters.

A second source of error is identified near the transmit
antenna. As characterized in the CST simulation
as shown in Fig.~\ref{fig:nearfield}, the Fraunhofer near-field boundary
of the BBHA~9120L antenna is $0.343 \text{m}$ at 3.349\,GHz. Positions~11 and~16, marked with $\times$ in
Fig.~\ref{fig:heatmaps}, are located at 0.367\,m and 0.351\,m
from the transmit antenna respectively, placing them within the
near-field to far-field transition zone. Sionna~RT models the
antenna using a far-field radiation pattern, which becomes less
accurate at distances close to $R_0$~. As
visible in the RSRP heatmap in Fig.~\ref{fig:rsrp}, Position~16 is located in the
forward beam direction and exhibits elevated prediction error,
consistent with the far-field approximation breaking down at
this distance. Position~11 is situated slightly behind the antenna
aperture in the back lobe region, shows comparatively lower
error as both the simulated and measured signal levels are weak,
reducing the absolute prediction difference.

Taken together, these observations demonstrate that accurate
system level RT validation in indoor environments requires both
precise electromagnetic material calibration of the digital
twin and careful consideration of antenna near-field boundaries
at measurement positions located close to the transmit
antenna.

\section{Conclusion}
\label{sec:conclusion}
This paper presented a hardware validated system level
evaluation of Sionna RT by directly comparing RT simulated
channels against VNA measured channels using an OAI based
5G NR testbed. By injecting both channel types into a
hardware in the loop channel emulator under identical network
conditions, KPI deviations in RSRP, PUCCH SNR, and SINR were quantified across 20 spatially distributed receiver
positions in an indoor laboratory environment.

The results identify two primary sources of RT prediction
error. First, material property mismatch in the digital twin
introduces localized multipath errors, particularly in regions
surrounding furniture with simplified material assignments,
highlighting the need for precise electromagnetic calibration
of indoor digital twins. Second, two measurement positions
located within the near-field to far-field transition zone of
the transmit antenna exhibit elevated prediction errors, as
the far-field radiation pattern assumption used by Sionna RT
becomes less accurate at distances close to the Fraunhofer
boundary $R_0 = 0.343$\,m, which was confirmed through
CST Microwave Studio simulation.

These findings provide quantitative insight into the
reliability of Sionna RT for system-level 5G performance
evaluation and establish a benchmark for digital twin-based
network planning. Future work will focus on refining material calibration and extending the evaluation to
larger indoor environments with higher Rx position density.

\balance
\bibliographystyle{IEEEtran}
\bibliography{/Users/faizanrauf/Desktop/Quantifying_System_Level_KPI_Deviations_Between_Sionna_Ray_Tracing_and_Measured_Channels_Using_OAI_Based_Evaluation/Ref.bib}

\end{document}